\documentclass[aps,prl,reprint,twocolumn,superscriptaddress,longbibliography,nofootinbib,floatfix,showpacs]{revtex4-1}
\usepackage{graphicx,color}
\usepackage{amsmath,amssymb}
\usepackage{hyperref}

\long\def\symbolfootnote[#1]#2{\begingroup%
\def\thefootnote{\fnsymbol{footnote}}\footnote[#1]{#2}\endgroup} 
\def\blfootnote{\xdef\@thefnmark{}\@footnotetext}

\definecolor{purp}{rgb}{0.5,0,0.5}
\definecolor{darkgreen}{rgb}{0.1,0.7,0}
\definecolor{orange}{rgb}{1,0.6,0}

\newcommand{\be}{\begin{equation}}
	\newcommand{\ee}{\end{equation}}
\newcommand{\bi}{\begin{itemize}}
	\newcommand{\ei}{\end{itemize}}
\newcommand{\bea}{\begin{eqnarray}}
	\newcommand{\eea}{\end{eqnarray}}

\newcommand{\bra}[1]{\langle\,#1\,|}          % Dirac bra
\newcommand{\ket}[1]{|\,#1\,\rangle}          % Dirac Ket

\begin{document}

\title{Quantum optical signatures in strong-field laser physics: Infrared photon counting in high-order-harmonic generation}

%% Notice placement of commas and superscripts and use of &
%% in the author list

\author{I. A.~Gonoskov}
\email[]{ivan.gonoskov@gmail.com}
\affiliation{Foundation for Research and Technology-Hellas, Institute of Electronic Structure $\&$ Laser, P.O. Box 1527,
GR-71110 Heraklion (Crete), Greece}
\affiliation{Max Planck Institute of Microstructure Physics, Weinberg 2, D-06120 Halle, Germany}

\author{{N.~Tsatrafyllis}}
\affiliation{Foundation for Research and Technology-Hellas, Institute of Electronic Structure $\&$ Laser, P.O. Box 1527,
GR-71110 Heraklion (Crete), Greece}
\affiliation{Department of Physics, University of Crete, 71103 Heraklion, Greece}
\author{I. K. Kominis}
\affiliation{Department of Physics, University of Crete, 71103 Heraklion, Greece}
\author{P.~Tzallas}
\email[]{ptzallas@iesl.forth.gr}
\affiliation{Foundation for Research and Technology-Hellas, Institute of Electronic Structure $\&$ Laser, P.O. Box 1527,
GR-71110 Heraklion (Crete), Greece}

\enlargethispage{10pt}
%%% 122 words
%++++++++++++++++++++++++++++++++++++++++++++++
%++++++++++++++++++++++++++++++++++++++++++++++
%++++++++++++++++++++++++++++++++++++++++++++++
\begin{abstract}
We analytically describe the strong-field light-electron interaction using a quantized coherent laser state with arbitrary photon number. We obtain a light-electron wave function which is a closed-form solution of the time-dependent Schr\"odinger equation (TDSE). This wave function provides information about the quantum optical features of the interaction not accessible by semi-classical theories. With this approach we can reveal the quantum optical properties of high harmonic generation (HHG) process in gases by measuring the photon statistics of the transmitted infrared (IR) laser radiation. This work can lead to novel experiments in high-resolution spectroscopy in extreme-ultraviolet (XUV) and attosecond science without the need to measure the XUV light, while it can pave the way for the development of intense non-classical light sources.\end{abstract}
\maketitle
\setlength{\belowdisplayskip}{0pt} \setlength{\belowdisplayshortskip}{0pt}
\setlength{\abovedisplayskip}{0pt} \setlength{\abovedisplayshortskip}{0pt}
%%% 122 words
%++++++++++++++++++++++++++++++++++++++++++++++
%++++++++++++++++++++++++++++++++++++++++++++++
%++++++++++++++++++++++++++++++++++++++++++++++
Strong-field physics and attosecond science \cite{Keldysh,reiss,Lew,Krausz} have been largely founded on the electron recollision process described by semi-classical approaches \cite{Lew} treating the electron quantum-mechanically and the electromagnetic field classically. This is because the high photon number limit pertinent to experiments with intense laser pulses appears to be adequately accounted for by a classically-described electromagnetic wave, which is not affected by the interaction.
%\vskip-0.6cm

In the semi-classical approaches (known as three-step models) used for the discription of the HHG process,  the electron tunnels through the strong-laser-field-distorted atomic potential, it accelerates in the continuum under the influence of the laser field and emits XUV radiation upon its recombination with the parent ion. Thus, the motion of the electron in an electromagnetic field is at the core of the recollision process. In the strong-field regime, this motion is well described by non-relativistic semi-classical Volkov wavefunctions, obtained by solving the TDSE for a free electron in a classically-described electromagnetic field. 
%\vskip-0.6cm

Extending the semi-classical Volkov wave functions into the quantum-optical region is non-trivial and, to our knowledge, a closed form solution of the quantized TDSE with a coherent-state light field has never been obtained before. Although an accurate calculation of the properties of the XUV radiation emitted from a gas phase medium requires the consideration of the driving IR laser bandwidth and the propagation effects in the atomic medium, the fundamental properties of the interaction can be adequately explored with the single-color single-atom interaction, as has been done in the pioneering work of Lewenstein et al.\cite{Lew}. In this work we develop a quantized-field approach for an ionized electron interacting with light field in a coherent state. We obtain a closed-form solution of the TDSE, which contains complete information about the laser-electron quantum dynamics during the interaction, and use it to describe the HHG process. Differently than previous approaches \cite{Gao,gaoguo1q2,Gao3,Gao4}, we 
describe the XUV emission as far-field dipole radiation by using an initially coherent laser state and the obtained closed-form electron-laser wave function, named "quantum-optical Volkov wave function". Our approach consistently extends the well-known semi-classical theories \cite{Lew}, since from the obtained quantum-optical wave function we can retrieve the semi-classical Volkov wave functions by averaging over the light states. This is of advantage, since all the results of the semi-classical theory (like harmonic spectrum, electron paths, ionization times, recombination times, etc.) can be retrieved from- and utilized in our quantized-field approach. 
%\vskip-0.6cm

Going beyond the reach of the semi-classical approach, we find that the quantum-optical properties of the HHG process are imprinted in measurable photon statistics of the transmitted IR laser field, thus accessing HHG dynamics does not require measuring the XUV radiation. This is a unique advantage of our work since our proposed measurements, dealing with  high-resolution spectroscopy in XUV and attosecond science, can be performed in open air without the need for specialized optics/diagnostics required for the characterization of the XUV radiation. Additionally, it has been found that the interaction of strong laser fields with gas phase media leads to the production of non-classical high photon number light states.\\
%\vskip-0.6cm
%++++++++++++++++++++++++++++++++++++++++++++++
%++++++++++++++++++++++++++++++++++++++++++++++
%++++++++++++++++++++++++++++++++++++++++++++++

\noindent\textbf{{Full-quantum theoretical description of light-electron interaction}}. 

The non-relativistic TDSE of an electron interacting with a single-mode long-wavelegth lineary-polarized quantized light field of frequency $\omega$ reads (in atomic units)

\begin{equation}\label{sch2}
	\begin{aligned}
		&i\frac{\partial{}\Psi}{\partial{t}}=\frac{1}{2}\Big[\,p\, - \hat{A} \Big]^{2}\Psi\,\,,
	\end{aligned}
\end{equation} 
where $p$ and $\hat{A}=-\frac{\beta}{\sqrt{2}}\big(\hat{a}e^{-i\omega{t}}+\hat{a}^{+}e^{i\omega{t}}\big)$ are the electron momentum and vector potential scalar operators along the polarization direction. The creation and annihilation operators are $\hat{a}^{+}=\frac{1}{\sqrt{2}}\left(q-\frac{\partial}{\partial{q}}\right)$ and $\hat{a}=\frac{1}{\sqrt{2}}\left(q+\frac{\partial}{\partial{q}}\right)$, respectively, $q$ is the in-phase quadrature of the field \cite{quant,quant2}, and $\beta=c\sqrt{2\pi/\omega V}$ is a constant determined by the quantization volume $V$, frequency $\omega$, and light velocity $c$. The detailed derivation of the analytical solution of Eq. (\ref{sch2}), termed quantum-optical Volkov wavefunction, will be given elsewhere. Here we provide the result, the validity of which can be checked by direct substitution into Eq. (\ref{sch2}). Based on this, we then analyze its fundamental features and their consequences for the HHG process. The closed-form solution of Eq. (\ref{sch2}) reads:

\begin{equation}\footnotesize
	\Psi(p,q,t)=C_{0}\sqrt{M(t)}\psi_{0}(p)\,e^{a(t)q^{2}+b(t)p^{2}+d(t)pq+f(t)p+g(t)q+c_{0}+c(t)}\label{sch4},
\end{equation}
where the functions $a(t), b(t),..., g(t), M(t)$ are given in terms of the parameters of Eq.(\ref{sch2}) in the Methods Section. The solution includes an arbitrary initial electron distribution $\psi_{0}(p)$ and an arbitrary initial photon number $N_{0}$ and the field phase $\theta$. The wavefunction $\Psi(p,q,t)$ provides the full quantum-optical description of the electron-light interaction. The term $d(t)pq$ in the exponent renders the electronic and light degrees of freedom non-separable. In the high photon number limit where $N_{0}\rightarrow\infty$, $\beta\rightarrow{0}$ ($V\rightarrow\infty$), and $\frac{\omega}{c}\beta{}\sqrt{2N_{0}}\rightarrow{}\frac{\omega}{c}A_{0}$ (where $A_{0}$ is the amplitude of the corresponding classically-described vector potential $A=A_{0}\cos{(\omega{t}+\theta}$)), Eqs.(\ref{sch4}) is simplified (see Methods) to ${\Psi}'(p,q,t)$ maintaining all the quantum-optical properties of Eqs.(\ref{sch2}).
%\vskip-0.6cm

A crucial  property of the quantum-optical Volkov wave function is that the matrix elements of any $q$-independent operator $\hat{R}$ coincide with the matrix elements obtained from using the well known semi-classical electron Volkov wave functions ${\psi_{V}}$ i.e.

\begin{equation}\label{i0i3}	
\bra{{\Psi_{x}}}\hat{R}\ket{{\Psi}'}\longrightarrow\bra{\psi_{x}}\hat{R}\ket{\psi_{V}},
\end{equation}
\vspace{0.01cm}

\noindent where $\Psi_{x}$ is an arbitrarily chosen electron-light wave function and $\psi_{x}$ is the corresponding state of the electron in case of classically-described electromagnetic field. Thus, while ${\Psi}'(p,q,t)$ goes beyond the semi-classical approach to completely describe the quantized electron and light interaction, it naturally reproduces the classical Volkov states after integrating over $q$. This has profound consequences for the description of HHG, since the well known results of the semi-classical models \cite{Lew} can be retrieved, and more than that, utilized in our quantized-field approach. In the recollision process, $\Psi_{x}$ is the ground state of the system ($\psi_{g}\psi_{c}$), $\hat{R}$ is the dipole moment ($\hat{r}$), $\psi_{x}$ is the ground state of the atom ($\psi_{g}$) and $\psi_{c}$ is the initial coherent light state. Detailed description of above considerations can be found in the Methods Section.
%\vskip-0.6cm

The calculation of the dipole moment in the high photon number limit demonstrates that the behavior of the electron in a strong laser field can be accurately described by semiclassical theories with negligible quantum corrections. However, our full quantum-optical approach can provide information about the IR laser field states during the interaction, inaccessible by the semi-classical theories. This information can be experimentally extracted using balanced homodyne detection \cite{homodyne, Mlynek, Bellini} of the IR laser field transmitted from the harmonic generation medium.\\
%\vskip-0.6cm

%++++++++++++++++++++++++++++++++++++++++++++++
%++++++++++++++++++++++++++++++++++++++++++++++
%++++++++++++++++++++++++++++++++++++++++++++++
\noindent\textbf{Quantum-optical description of the HHG process}. 

Using the quantum-optical Volkov wave functions $\Psi'$, the time evolution of the HHG process is described by the following wave function

\begin{equation}\label{i0i1}
	\tilde{\Psi} = a_{g}\Psi_{g} + \sum\limits_{i}b_{i}\Psi'(p,q,t-t_{i})\;,
\end{equation}
where $\Psi_{g}=\psi_{g}\psi_{c}$ is the initial state of the system, $\psi_{g}$ is the ground state of the electron, $\psi_{c}$ is the initial coherent light state and $\Psi'(p,q,t-t_{i})$ are the continuum laser-electron states having different ionization times $t_{i}$. The complex amplitudes $a_{g}$ and $b_{i}$ satisfy the normalization condition $\langle\tilde{\Psi}|\tilde{\Psi}\rangle=1$. In this case, the time dependent dipole moment is $r(t)=\bra{\tilde\Psi}\hat{r}\ket{\tilde\Psi}=\int_{-\infty}^{\infty}\int_{-\infty}^{\infty}\tilde{\Psi}\hat{r}\tilde{\Psi}^{*}dqdp$. As in the semi-classial theory \cite{Lew}, we neglect ground-ground $\bra{\Psi_{g}}\hat{r}\ket{\Psi_{g}}$ and continuum-continuum $\bra{\Psi'}\hat{r}\ket{\Psi'}$ transitions, and consider only ground-to-continuum (and continuum-to-ground) transitions given by the matrix elements $\bra{\Psi'}\hat{r}\ket{\Psi_{g}}$ (and $\bra{\Psi_{g}}\hat{r}\ket{\Psi'}$). 
%\vskip-0.6cm

Integrating $\langle\tilde{\Psi}|\hat{r}|\tilde{\Psi}\rangle$ over $q$, using Eq. (4) and integrating over $p$, we retrieve that $r(t)\propto\left\langle {\psi}_{V}\big|\,{\hat{r}\,}\big|\psi_{g}\right\rangle$ which coincides with the expression given by the semi-classical theories. Thus, all the semi-classical results \cite{Lew,bell,corsi,zair}, in particular the short (S) and long (L) electron paths with electronic Volkov wave functions ${\psi}_{V}^{S}$ and ${\psi}_{V}^{L}$ respectively, can be consistently used in the present approach. In a similar way, the same results can be obtained using the IR wave functions $\psi_{l}$ (see Method section).  
%\vskip-0.6cm

A scheme which can describe the HHG process in the context of the present model is shown in Fig.1. Although the quantization of the harmonic field is not required for this work and thus was not considered in the previous formalism, harmonic photons are included in Fig. 1 for a complete understanding of the process. Fig.1a shows the electron states in case of integrating over $q$ and Fig.1b shows the field states in case of integrating over $p$. The horizontal black lines in Fig.1a and 1b are the initial states of the electron $|\psi_g\rangle$ and IR-laser field $|\psi_c\rangle$ with energy $IP<0$ and $W_{\rm light}(0)$, respectively. At $t>0$ the system is excited (small red arrows) in an infinite number of entangled laser-electron $\Psi'_{i}$ states (gray area), resulting in a reduction of the average laser energy (small downwards red arrows in Fig. 1b) and the enhancement of the average electron energy (small upwards red arrows in Fig. 1a). The $\Omega$-frequency emission is taking place by constructive 
interference of $\psi_{V}$ states and recombination to the ground state (downwards red arrows in Fig.1a). In this case the final laser energy remains shifted by $\hbar\Omega$ compared to the initial energy $W_{\rm light}(0)$. When the $\psi_{V}$ states interfere destructively the probability of $\Omega$-frequency emission is reduced and the average laser energy returns to the initial value (black dashed arrows in Fig.1b). Among the infinite $\Psi'_{i}$ states, two are the dominant surviving the superposition, corresponding to the S and L electron paths, described by $\ket{\psi_{V,\Omega}^{\rm S}}$ and $\ket{\psi_{V,\Omega}^{\rm L}}$, respectively. Correlated to them are the IR-laser states $\ket{\psi_{l,\Omega}^{\rm S}}$ and $\ket{\psi_{l,\Omega}^{\rm L}}$ (where ${\psi}_{l}^{S}$ and ${\psi}_{l}^{L}$ are the IR wave functions correspond to the S and L electron paths, respectively) , as well as the $\Omega$-frequency states $\ket{\psi_{\Omega}^{\rm S}}$ and $\ket{\psi_{\Omega}^{\rm L}}$, respectively. This is 
consistent with the interpretation of recent experimental data \cite{Kom}.
\begin{figure}
	\includegraphics[width=\columnwidth]{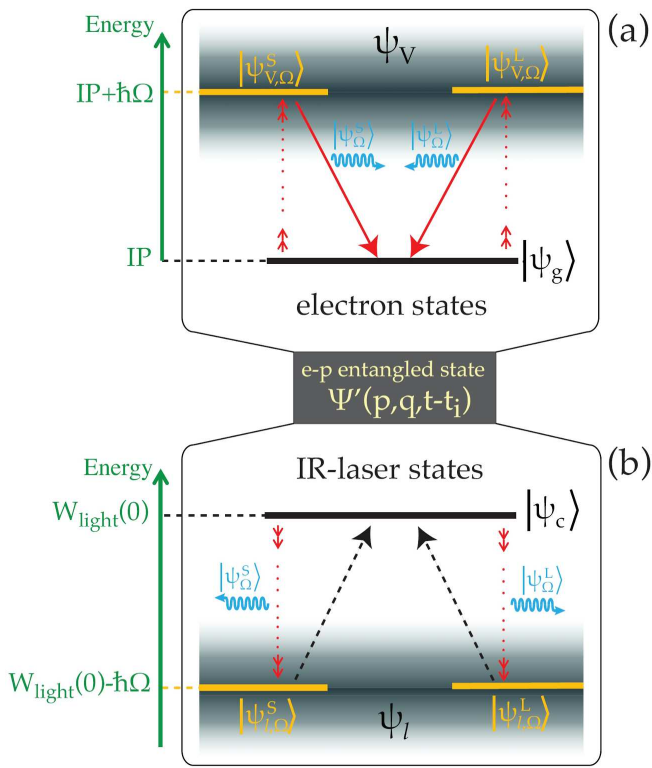}
	\caption{Excitation scheme for the quantum-optical description of the HHG process. (a) A schematic representation of the electron states in case of integrating over laser-state parameter $q$. (b) A schematic representation the laser field states in case of integrating over electron momentum $p$.} 
\end{figure}
It is thus evident that by measuring quantum optical properties of the IR-light we can access the full quantum dynamics of the HHG process. Such properties, in particular photon statistics to be discussed next, are not accessible by semi-classical models \cite{Keldysh,reiss,Lew,Krausz}.\\
%\vskip-0.6cm

%++++++++++++++++++++++++++++++++++++++++++++++
%++++++++++++++++++++++++++++++++++++++++++++++
%++++++++++++++++++++++++++++++++++++++++++++++
\noindent\textbf{Counting IR photons in HHG}. 

The probability for measuring $n$ photons in a non-interacting coherent light state is given by $P_{n}=\left|K_{n}\right|^{2}$, where $K_{n}$ is a probability amplitude appearing in the expansion $\Psi=\sum\limits_{n=0}^{\infty}K_{n}\,\ket{n}$, in terms of photon-number (Fock) states. In this expression, $K_{n}$ is time-independent, since, as well known \cite{knight}, the photon probability distribution in a coherent state is constant within the cycle of the light field (Fig. 2a). When the coherent light state is interacting with a single atom towards the generation of XUV radiation, the probability distribution becomes time dependent, since $\Psi'(p,q,t-t_{i})$ is changing at each moment of time within the cycle of the laser field due the interaction with the ionized electron. In this case, the probability distribution is given by $P_{n} (t)=\left|\sum\limits_{\Omega_{i}}\sum\limits_{k_{\Omega_{i}}}{K}_{n}^{(k_{\Omega_{i}},\Omega_{i})}(p_{i}^{(\Omega_{i})}(t),t, 
t_{i}^{(\Omega_{i})}, t_{r}^{(\Omega_{i})})\right|^{2}$, where $t_{i}^{(\Omega_{i})}$ and $t_{r}^{(\Omega_{i})}$ are the ionization and recombination times of the corresponding electron paths $k_{\Omega_{i}}=S_{\Omega_{i}},L_{\Omega_{i}}$ with momentum $p_{i}^{(\Omega_{i})}(t)$ which lead to the emission of XUV radiation with frequency $\Omega_{i}$ (see Methods Section). The parameters $t_{i}^{(\Omega_{i})}$, $t_{r}^{(\Omega_{i})}$ and $p_{i}^{(\Omega_{i})}(t)$ are obtained using the 3-step semi-classical model \cite{Lew}.
%\vskip-0.6cm	

In reality, an intense Ti:S femtoseond (fs) laser pulse with $\sim{}10^{17}$ photons/pulse (which corresponds to $N_{0}\sim{}10^{12}$ photons/mode for a laser system based on a $100$ MHz oscillator which delivers pulses of $\sim{}30$ fs duration), interacts with gas-phase medium towards the emission of XUV radiation. In this case, where $n_a^{(Q_{i})}$ atoms coherently emit XUV radiation with frequencies proportional ($Q_{i}=\Omega_i/\omega$) to the frequency of the IR laser, the interaction is imprinted in the photon number $\bar{N}$ of the IR field as $\bar{N}=N_{0}-n_a^{(Q_{i})}Q_{i}$, reflecting energy conservation. Since the signal of interest, $n_a^{(Q_{i})}Q_{i}$, is superimposed on a large background ($\approx{}N_0$), a balanced interferometer \cite{homodyne, Mlynek, Bellini} is required in order to subtract the initial IR photon number $N_{0}$ from $\bar{N}$ and thus measure $\Delta{N}=N_0-\bar{N}=n_{a}^{(Q_{i})}Q_{i}=C^{(Q_{i})}n_{\rm int}Q_{i}$. The number of atoms interacting with the laser field 
is $n_{\rm int}$ and $C^{(Q_{i})}$ is the conversion efficiency of a single-XUV-mode, which  
depends on the gas density in the interaction region. Taking into account that for gas densities $\sim{}10^{18}$ atoms/${\rm cm}^{3}$ the conversion efficiency is $\sim{}10^{-4}$ (for Argon, Krypton, Xenon in the 25-eV photon energy range) \cite{Hergott, Tzallas1, Skantzakis, Tzallas2, Mido}, it can be estimated that $\Delta{}N$ ranges from $\sim{}0$ (for zero gas density) up to $\sim{}10^{9}$ photons/mode (for gas density $\sim{}10^{18}$ atoms/${\rm cm}^{3}$). Although the study is valid for all noble gases, in the following we will describe the HHG process considering Xenon atoms interacting with a coherent IR laser field in case of low ($n_a^{(Q_{i})}=1$), intermediate ($n_a^{(Q_{i})}=100, 500$) and high ($n_a^{(Q_{i})}=10^8$) number of emitting atoms.
\begin{figure*}
	\includegraphics[width=\textwidth]{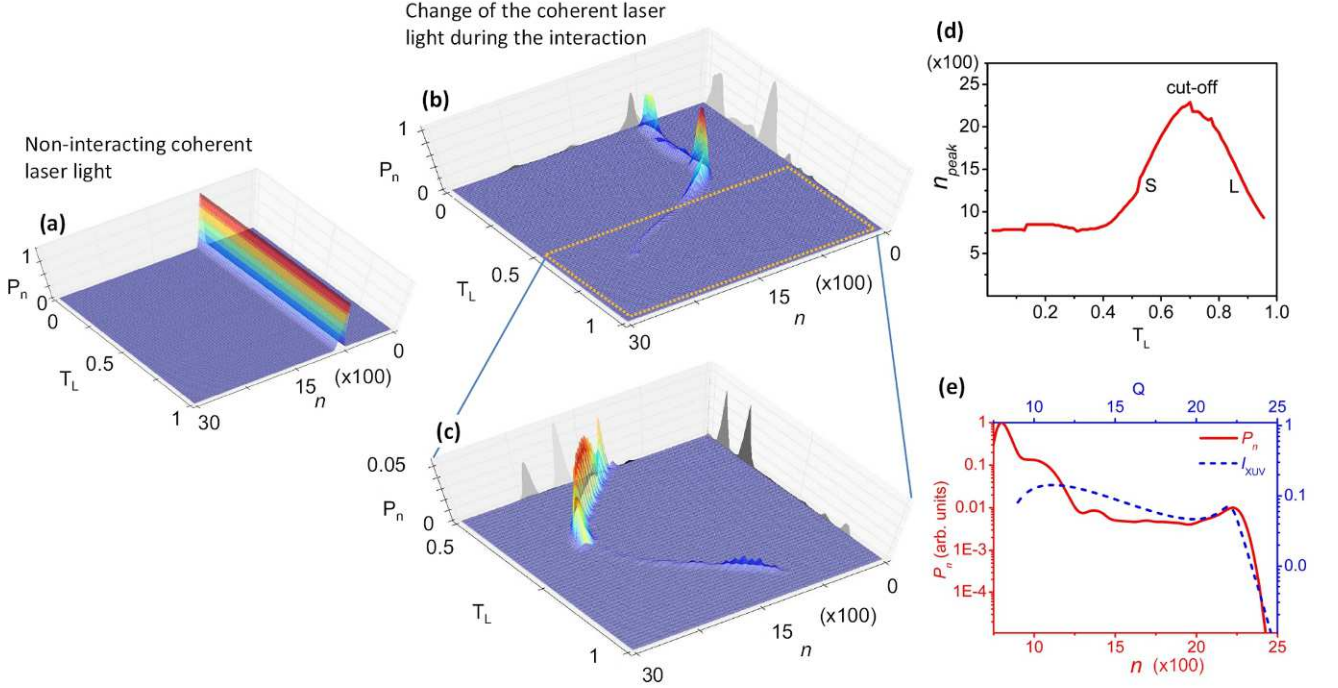}
	\caption{Probability distribution of the IR photons during the HHG process. (a) Probability distribution of a non-interacting coherent IR laser state for $N_0=800$. This is shown only for reasons of comparison with the interacting coherent laser states. (b) Time dependence of the IR probability distribution during the recollision process calculated for $n_{a}^{(Q_{i})}=100$ and $I_l=10^{14}W/cm^2$. In the calculation, the electron momentum, the ionization and the recombination times have been obtained by the 3-step semi-classical model. (c) Expanded plot of Fig. 2b in the time interval $0.5 T_L < t < T_L$ (d) IR photon number absorbed by Xenon atoms during the recollision process. This has been obtained by the $n=n_{peak}$ position of the peak of the distribution at each moment of time. (e) Overall IR photon number absorbed by the atoms during the recollision (red solid line). This has been obtained after integration over the cycle of the IR field. The XUV spectrum shown in blue dashed line, 
obtained using the semi-classical 3-step model.} 
\end{figure*} 
%\vskip-0.6cm	
\^{e}
For a single recollision, the dependence of $P_{n} (t)$ on time during the process is shown in Figs. 2b,c  for $n_{a}^{(Q_{i})}=100$. It is seen that in the time interval $0<t<t_{i}\approx300$ asec where the ionization is taking place, the peak of the probability distribution is located at $n=n_{peak}\approx800$. Since the ionization of one Xenon atom requires the absorption of $n\approx8$ IR photons, this value corresponds to the energy absorbed by 100 Xenon atoms.  For $t>t_{i}$ the variation of the IR photon number $n$ with time reflects the energy exchange between the IR laser field and the free electron. The peak of the probability distribution during the recollision is located at $n=n_{peak}=\Omega_{i}/\omega$, $t=t^{(\Omega_{i})}_{r}$ (Fig. 2d). This is due to the energy absorbed by Xenon atoms during the recollision process towards the emission of XUV radiation with frequency $\Omega_{i}$ at the moment of recombination $t^{(\Omega_{i})}_{r}$. Importantly, in Fig. 2d we demonstrate that the absorbed 
IR photon number reveals the fundamental properties of the three-step semi-classical model: S and L paths lead to the emission of the same XUV frequency and degenerate to a single path in the cut-off region. Furthermore, as shown in Fig. 2e, the overall IR photon number distribution (red solid line) reproduces the well-known XUV spectrum resulting from the semi-classical three-step model (blue dashed line), including the  plateau and cut-off regions. Thus, we demonstrated that all known features of the semiclassical three-step model are imprinted in IR photon statistics. 
%\vskip-0.6cm

We will now explore the new phenomena and potential metrological applications one can address utilizing IR photon statistics. To this end, we first elaborate on the atom-number dependence of $P_n$. While the number of IR photons absorbed by the system is proportional to $n_{a}^{(Q_{i})}$, the width $w_{n_{a}}(t)$ of the probability distribution is determined by Gaussian statistics, $w_{n_{a}}(t)\propto\sqrt{n_{a}^{(Q_{i})}}$ (Figs 3a). However, the distribution is departing from the Gaussian statistics during the recollision process. This is clearly shown in Fig. 3b which depicts in contour plot of the normalized probability distribution of Fig 3a. For reasons of comparison, a Gaussian distribution is shown in Fig. 3c. The distortion of the probability distribution in Fig. 3b, more pronounced in the time interval $0.5 T_L < t < T_L$, is associated with energy/phase dispersion of the interfering electron wave packets in the continuum, alluding the possibilities of producing non-classical light-states.
%\vskip-0.6cm

For multi-cycle laser field, the process is repeated every half-cycle of the laser period. In this case the probability distribution consists a series of well confined peaks (Fig. 4a,b) appearing at positions $n=\tilde{Q}=\Omega/\omega$ and reflects the formation of well confined high order harmonics ($\tilde{Q}$). 
%\vskip-0.6cm

\begin{figure}
	\includegraphics[width=\columnwidth]{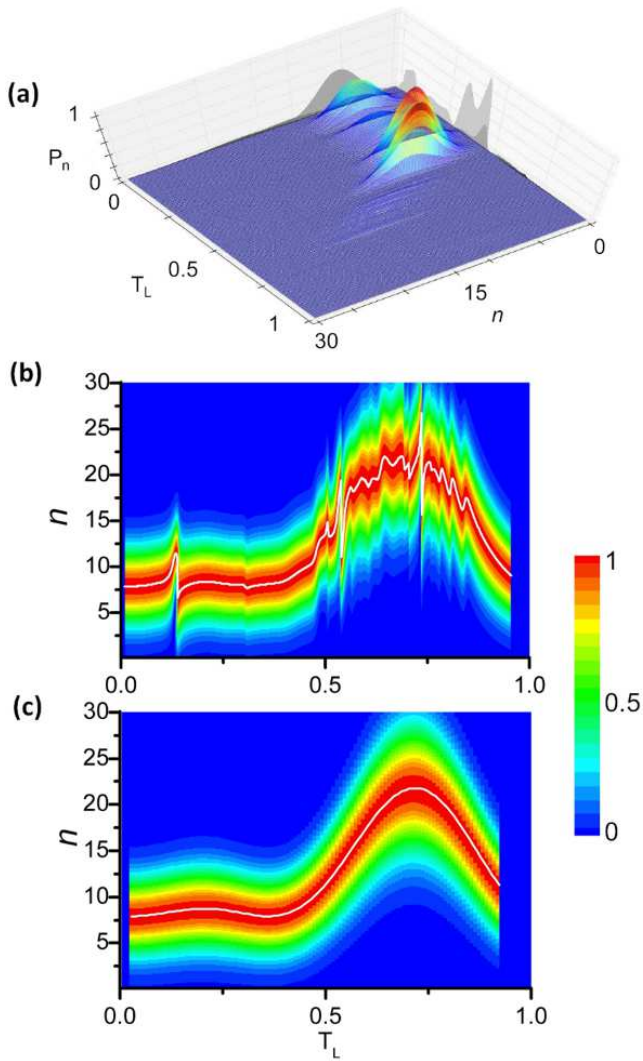}
	\caption{Generation of light states with non-Gaussian photon distribution. (a) Time dependence of the IR probability distribution during the recollision process calculated $n_{a}^{(Q_{i})}=1$ and $I_l=10^{14}W/cm^2$. In the calculation, the electron momentum, the ionization and the recombination times have been obtained by the 3-step semi-classical model. (b) Contour plot of the normalized probability distribution of Fig. 3a. (c) Contour plot of the normalized probability distribution which follows the Gaussian photon statistics. This has been calculated using a single electron path which contributes to the emission of a monochromatic XUV radiation with frequency $Q_{i}=\omega/11$. It is evident, that in case of reducing the number of emitting atoms form $n_{a}^{(Q_{i})}=100$ (Fig. 2b,c) to $n_{a}^{(Q_{i})}=1$ the width of the probability distribution is increasing.} 
\end{figure}
Additionally, the atom-number dependence of the IR photon distribution in the HHG process provides significant advantages for high resolution spectroscopy in XUV and attosecond science. In Fig. 4c (left panel) we show the dependence of the $P_{n}$ on the intensity of the laser field ($I_l=\varepsilon_{0}|E_{0}|^2/2\propto N_{0}$) and $n$ for $n_{a}^{(\tilde{Q})}=10^8$ (for simplicity we consider only the case of $\tilde{Q}=11,\;13,\;15$ ). Indeed, the harmonic spectrum can be obtained from the maxima of $P_n$ centered at $\Delta{}N=\tilde{Q}\cdot{}n_{a}^{(\tilde{Q})}$. The spacing between the maxima $\delta(\Delta{N})=n_{a}^{(\tilde{Q})}\delta{\tilde{Q}}$ with $\delta{\tilde{Q}}$=2 for consecutive harmonics, and the width $w=\sqrt{2\Delta{}N}=\sqrt{2n_{a}^{(\tilde{Q})}\tilde{Q}}$ depend on $C^{(\tilde{Q})}$ and $n_{\rm int}$. The resolving spectral power $P_{R}=\Delta{}N{}/w=\sqrt{\Delta{}N/2}=\lambda_{\Omega}/\delta\lambda_{\Omega}$ increases with $n_{\rm int}$ and for values of $\Delta{}N\sim{}10^{9}$ 
photons, $P_R$ can reach the values of $\sim{} 10^{4}-10^{5}$ in the spectral range of 25 eV, which competes with state-of-the-art XUV spectrometers. This is shown in Fig. 4c (right panel) where the probability distribution around $\tilde{Q}=15$ has been calculated in case of recording the 799.95nm, 800.00 nm and 800.05 nm IR modes of a Ti:S laser pulse. This measurement can be performed by collecting the photons of the IR modes of the spectrally resolved multi-color IR pulse. This can be done by means of an IR diffraction grating placed after the harmonic generation medium. This figure also depicts the broadening effects introduced in a measured distribution by the bandwith of the driving IR pulse in case of collecting more than one modes of the multi-mode laser pulse.
%\vskip-0.6cm

\begin{figure*}
	\includegraphics[width=12cm]{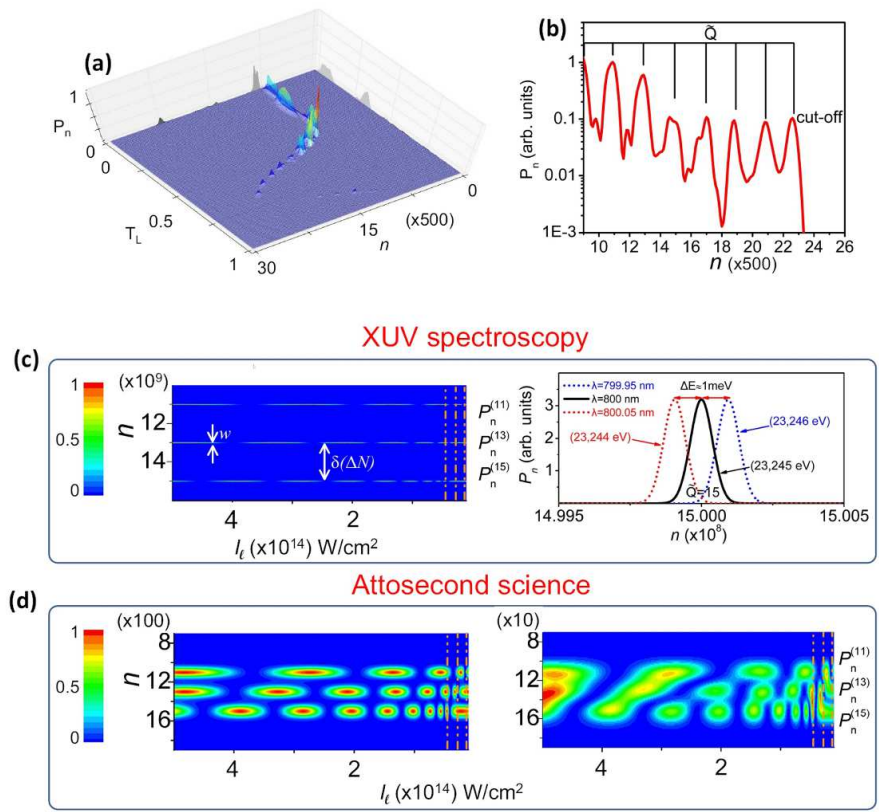}
	\caption{High resolution spectroscopy in XUV and attosecond science using IR photon statistics. (a) Probability distribution for multi-cycle laser interaction calculated for $n_{a}^{(Q_{i})}=500$ and $I_l=10^{14}W/cm^2$. For this graph three laser cycles have been considered. (b) "IR photon statistics spectrum" obtained by time integrating the Fig. 4a. (c) (left panel) Dependence of $P_{n}$ on the laser intensity $I_l$ ($\propto{}N_{0}$) and on photon number $n$ for $\tilde{Q}=11,\;13,\;15$ and $n_{a}^{(Q_{i})}=10^{8}$. The right panel shows the probability distribution around $\tilde{Q}=15$ in case of recording the 799.95nm, 800.00 nm and 800.05 nm IR modes of the laser pulse after passing through the gas medium. (d) Dependence of $P_{n}$ on the laser intensity and on photon number $n$ for $\tilde{Q}=11,\;13,\;15$ and for $n_{a}^{(Q_{i})}=100$ (left panel) and $n_{a}^{(Q_{i})}=10$ (right panel). The dashed vertical lines depict the cut-off positions of the harmonics. In these plots $n_{a}^{(Q_{i})}
$ were taken independent of $I_l$.} 
\end{figure*}

When $\Delta{N}$ is reduced, the probability distribution is getting broader (Fig. 4d, left panel), while at the point where the probability distribution between the consecutive harmonics overlaps, an interference pattern associated with the relative phase between the consecutive harmonics appears in Fig. 4d (right panel). Additionally, the modulation of $P_{n}$ with the intensity of the laser field (clearly shown in the left panels of Figs. 4c,d) reflects the effect of the S and L path interferences in the context of Fig. 1, i.e. the maxima (minima) of $P_{n}$ versus $N_{0}$ correspond to those IR-laser intensities $N_{0}$, for which $\Psi_{V}$ interferes destructively (constructively). These observations can be used for attosecond science and metrology, to be explored in detail elsewhere. Since the photon statistics measurements are sensitive to shot-to-shot fluctuations of the IR intensity, stable laser systems or IR energy tagging approaches are required in order to be able to record an "IR photon 
statistics spectrum". Additionally, in order to avoid the influence of the laser intensity variation along the propagation axis in the harmonic generation medium, a gas medium with length much smaller compared to the confocal parameter of the laser beam is required. Any influence of the intensity variation along the beam profile at the focus can be minimized (in case that is needed) using spatial filtering approaches where the IR photons of the specific area on the focal spot diameter can be collected.     
%++++++++++++++++++++++++++++++++++++++++++++++
%++++++++++++++++++++++++++++++++++++++++++++++
%++++++++++++++++++++++++++++++++++++++++++++++
\section*{Conclusions}
%\vskip-0.6cm

Concluding, we have developed a quantized-field approach which describes the strong-field light-electron interactions using a quantized coherent laser state with arbitrary photon number. The description is based on the quantized-Volkov light-electron wave function resulting from the closed-form solution of TDSE. The obtained wave function provides information about the quantum optical features of the interaction, which are not accessible by the semi-classical approaches used so far in strong-field physics and attosecond science. The approach has been used for the description of HHG in gases. We have found that the quantum optical features of the HHG can be unraveled by measuring the photon statistics of the IR laser beam transmitted from the gas medium without the need of measuring the XUV radiation. This is a unique advantage of the work since our proposed measurements, dealing with high-resolution spectroscopy in XUV and attosecond science, can be performed without the need for specialized XUV equipment (
gratings, mirrors, high vacuum conditions e.t.c.). Additionally, we have found that the HHG process in gases can lead to non-classical IR light states. In general, this work establishes a promising connection of strong-field physics with quantum optics.

%++++++++++++++++++++++++++++++++++++++++++++++
%++++++++++++++++++++++++++++++++++++++++++++++
%++++++++++++++++++++++++++++++++++++++++++++++
\section*{Methods}
%\vskip-0.6cm

\noindent\textbf{On the closed-form solution of TDSE}: In order to obtain a closed-form solution of Eq. (\ref{sch2}) of the main text of the manuscript, we consider as an initial state, a state where the electron is decoupled from the light i.e. $\Psi_{0}=\Psi(p,q,t=0)=\psi_{c}(q)\psi_{0}(p)$ to be a separable product of a coherent state of light $\psi_{c}(q)=e^{\lambda{}q^{2}+g_{0}q+c_{0}}$ and an arbitrary field-independent electron state $\psi_{0}(p)$ in momentum representation, where $\lambda{}=-\frac{1}{2}\sqrt{1+\frac{\beta^{2}}{\omega}}$ is the parameter which introduces the light dispersion due to the presence of the electron \cite{varro1q,my1q}, $g_{0}=\sqrt{2N_{0}}e^{-i\theta}$ carries the information about the phase of the light ${\theta}$, $N_{0}$ is the average photon number of the initial ($t=0$) coherent light state, and $c_0$ is a normalizing constant.\linebreak
\noindent The parameters appearing in Eq. \eqref{sch4} of the main text of the manuscript are 

\begin{align}\nonumber
	a(t)&=-\frac{1}{2}+\left(\frac{1}{2}+\lambda\right)M\,e^{2i\omega{t}},\\ \nonumber
	b(t)&=-it\left(\frac{1}{2}-\frac{\beta^{2}}{8\lambda^{2}\omega}\right)+\frac{\beta^{2}}{8\lambda^{3}\omega^{2}}(1-e^{2i\omega{t}}) \\ 	
            &~~~+\frac{\beta^{2}}{32\lambda^{3}\omega^{2}}(1-e^{4i\omega{t}})+\frac{\gamma^{2}m}{2(1-2\lambda)},\nonumber\\ \nonumber
	d(t)&=\gamma\,M\,e^{i\omega{t}},\\ \nonumber
	f(t)&=\frac{g_{0}\gamma{}e^{2i\lambda\omega{t}}}{(1-2\lambda)}m-\frac{g_{0}\omega\gamma^2}{2\beta}, \\
	g(t)&=g_{0}\,M\,e^{2i\lambda\omega{t}}e^{i\omega{t}},\nonumber \\
	c(t)&=i\omega{t}\left(\frac{1}{2}+\lambda\right)-\frac{g_{0}^{2}}{8\lambda}(1-e^{4i\lambda\omega{t}})\nonumber+\frac{g_{0}^{2}e^{4i\lambda\omega{t}}}{2(1-2\lambda)}m\nonumber
\end{align}
where $\gamma=\frac{\beta}{2\lambda\omega}(1-e^{2i\lambda\omega{t}})$, $M=\left[(\frac{1}{2}-\lambda)+(\frac{1}{2}+\lambda)e^{2i\omega{t}}\right]^{-1}$, $m=1-Me^{2i\omega{t}}$, $g_{0}=\sqrt{2N_{0}}e^{-i\theta}$, $N_{0}=\left\langle\psi_{c}\left|\hat{a}^{+}\hat{a}\right|\psi_{c}\right\rangle$ and ${C}_{0}$ is normalization constant. From the general solution (\ref{sch4}) we can recover energy conservation, i.e. the instantaneous interaction energy of the electron is given by $W_{\text{e,int}}(t)=W_{e}(t)-W_{e}(0)=W_{\text{light}}(0)-W_{\text{light}}(t)$, where $W_{e}(0)$ is the initial kinetic energy of the electron, $W_{\text{light}}(0)=\omega\sqrt{1+\frac{\beta^{2}}{\omega}}\Big[\frac{1}{2}+N_{0}\Big]$ is the initial energy of the light field, $W_{\rm light}(t)=\omega\sqrt{1+\frac{\beta^{2}}{\omega}}\Big[\frac{1}{2}+\bar{N}(t)\Big]$ is the field energy at any moment of time and $\bar{N}(t)=\left\langle N \right\rangle = \left\langle\Psi\left| \hat{a}^{+}\hat{a} \right|\Psi\right\rangle$. In the high photon 
number limit the $q$-dependent part of the total wave function becomes exponentially small everywhere except the region around $\left|q\right|\approx\sqrt{2N_{0}}$. Thus, Eq. (\ref{sch4}) of the main text of the manuscript leads to

\begin{align}
	{\Psi}'(p,q,t)=C_{0}'\psi_{0}(p)e^{a'(t)\,q^{2}+b'(t)\,p^{2}+d'(t)\,pq+f'(t)\,p+g'(t)\,q}\nonumber,
\end{align}
where now the parameters in the exponent are
\begin{align}\nonumber
	a'(t)&=-\frac{1}{2}-\frac{1}{4}\frac{\beta^{2}}{\omega}e^{2i\omega{t}}+O(\beta^{4}),\\\nonumber
	b'(t)&=-it\frac{1}{2}+O(\beta^{2}),\\\nonumber
	d'(t)&=-\frac{\beta}{\omega}(e^{i\omega{t}}-1)+O(\beta^{2})\nonumber\\\nonumber
	f'(t)&=-\frac{1}{\omega}A_{0}e^{-i\theta}(1-\cos{\omega{t}})+O(\beta),\\\nonumber
	g'(t)&=\sqrt{2N_{0}}\,e^{-i\left(\theta+\frac{\beta^{2}}{2}t\right)}\bigg[1-\frac{1}{4}\frac{\beta^{2}}{\omega}(1-e^{2i\omega{t}})\bigg]+O(\beta^{3})\nonumber
\end{align}
and ${C}_{0}^{'}$ is normalization constant.\\

%++++++++++++++++++++++++++++++++++++++++++++++
%++++++++++++++++++++++++++++++++++++++++++++++
%++++++++++++++++++++++++++++++++++++++++++++++
%\vskip-0.6cm
\noindent\textbf{On the validity of Eq. (\ref{i0i3})}: Eq. (\ref{i0i3}) of the main text of the manuscript can be rigorously proved, since the integration over $q$, $\int\limits_{-\infty}^{\infty}\int\limits_{-\infty}^{\infty}\,{\Psi_{x}}{\hat{R}}{\Psi}'^{*}\,dp\,dq$, leads to the delta function $\delta\left(q-\sqrt{2N_{0}}\cos{\theta}\right)$, and results in integral over $p$ which is equal to $\bra{\psi_{x}}\hat{R}\ket{\Psi_{V}}$.\\
%++++++++++++++++++++++++++++++++++++++++++++++
%++++++++++++++++++++++++++++++++++++++++++++++
%++++++++++++++++++++++++++++++++++++++++++++++
%\vskip-0.6cm
\noindent\textbf{On the description of HHG using the IR wave functions}: The results obtained by the semi-classical theories regarding HHG can be also obtained by integrating $\langle\tilde{\Psi}|\hat{r}|\tilde{\Psi}\rangle$ over $p$, using Eq. (\ref{i0i3}) and integrating over $q$. In this case the dipole moment $r(t)\propto\left\langle {\psi}_{l}^{V}\big|\psi_{c}\right\rangle$ is expressed in terms of the corresponding to the Volkov-electron states IR wave functions ${\psi}_{l}^{V}=\int{\Psi'(p,q,t-t_{V})}\,\nabla_{p}\,{\psi}^{*}_{g}(p)\,dp$ (where ${\psi}_{g}(p)$ is a ground state of the electron in momentum representation, $t_{V}$ are the ionization times Volkov electron paths contribute to the harmonic generation). The IR wave functions which correspond to the S and L electron paths are $\psi_{l}^{S}=\int{\Psi'(p,q,t-t_{S})}\,\nabla_{p}\,{\psi}^{*}_{g}(p)\,dp$ and $\psi_{l}^{L}=\int{\Psi'(p,q,t-t_{L})}\,\nabla_{p}\,{\psi}^{*}_{g}(p)\,dp$, respectively, with $t_{S}$ and $t_{L}$ being the ionization times 
of the short and long electron paths.\\
%++++++++++++++++++++++++++++++++++++++++++++++
%++++++++++++++++++++++++++++++++++++++++++++++
%++++++++++++++++++++++++++++++++++++++++++++++
%\vskip-0.6cm
\noindent\textbf{On the calculations of the IR probability distribution}: The probability to measure $n$ photons in a non-interacting light field state $\Psi$ is $P_{n}=\left|K_{n}\right|^{2}$, where $K_{n}$ is a probability amplitude appearing in the expansion $\Psi=\sum\limits_{n=0}^{\infty}K_{n}\,\ket{n}$, in terms of photon-number (Fock) states. In $q$-representation the Fock states are written as $\ket{n}=\frac{\exp\left(-q^{2}/2\right)}{\pi^{1/4}2^{n/2}\sqrt{n!}}H_{n}(q)$, where $H_{n}(q)=2^{n}\exp[{-\frac{1}{4}\frac{\partial^{2}}{\partial{q}^{2}}}]\;q^{n}$ are Hermite polynomials. For coherent light states \cite{quant}, the photon statistics are described by the Poisson distribution $P_{n}=\frac{N_{0}^{n}}{n!}e^{-N_{0}}$, well-approximated by a Gaussian $P_{n}\approx\frac{1}{\sqrt{2\pi{}N_{0}}}\,\exp\big[{-\frac{(n-N_{0})^{2}}{2N_{0}}}\big]$ when $N_{0}\gg{1}$. 
In case of HHG process, the probability distribution during the recollision process for a single path $i$ of ionization time $t_{i}$ and electron momentum $p_{i}(t)$ which contributes to the production of XUV radiation with frequency the harmonic $\Omega_{i}$ is given by $P_{n}=\left|K_{n}^{(i)}(t)\right|^{2}$, where $K_{n}^{i}(t)\approx c_{i}(t)\frac{1}{\sqrt{n!}}\bigg[\sqrt{N_{0}}e^{-i\theta_{i}}+\frac{A_{0}\,p_{i}(t)}{2\omega\sqrt{N_{0}}}\left(1-e^{i\omega({t-t_{r}})}\right)\bigg]^{n}=c_{i}(t)\tilde{K}_{n}^{i}(t)$ is determined through the expansion $\Psi'(p_{i}(t),q,t_{i})=\sum\limits_{n=0}^{\infty}K_{n}^{i}(t)\,\ket{n}$. $c_{i}(t)$ are $n$-independent complex numbers proportional to the $q$-independent part of the $\Psi'$ and $\theta_{i}=\omega(t-t_{i})$ is the phase of the laser field at the moment of ionization. In the high photon number limit, $c_{i}(t)\tilde{K}_{n}^{i}(t)\rightarrow{}A_{i}(t)\,e^{i\Phi_{i}}$ (where $A_{i}(t)$ is real), and the probability distribution reads $P_{n}=\left|A_{i}(t)\,e^{
i\Phi_{i}}\right|^{2}$, with  
\begin{equation}\label{i0i5p}\small
	\begin{aligned}
		\Phi_{i}\approx-(t-t_{r})\frac{[p_{i}(t)]^{2}}{2}+\frac{p^{i}(t)A_{0}}{\omega}\sin(\theta_{i}^{(1,2)})\left[1-\cos(\omega{}(t-t_{r}))\right] \\ -\frac{p_{i}(t)A_{0}}{\omega}\,\frac{n}{N_{0}}\cos(\theta_{i})\sin(\omega{}(t-t_{r}))\nonumber
	\end{aligned}
\end{equation} 
where $\left|{A}_{n}(t)\right|^{2}\propto\exp\left[-({n-\bar{N}(t))^{2}}/{2\bar{N}(t)}\right]$ and $\bar{N}(t)=N_{0}-\xi(t)$ is the average number of photons during the recollision, with $\xi(t)=\Omega(t)/\omega$ and $\Omega(t)=((p^{2}(t)/2)-IP)$. When multiple paths contribute to the emission of multiple harmonics, $P_{n}\approx\left|\sum\limits_{\Omega_{i}}\sum\limits_{k_{\Omega_i}}{A}_{n}^{(k_{\Omega_i},\Omega_i)}\,e^{i\Phi_{i}}\right|^{2}$,
where $k=S_{\Omega_i},L_{\Omega_i}$ denotes the electron paths contribute to the emission of the $\Omega_{i}$ frequency. Since the probability distribution during the recollision is located at ($n=n_{peak}=\Omega_{i}/\omega$, $t=t^{(\Omega_{i})}_{r}$) the above expression of $P_{n}$ and $\Phi_i$ can be further simplified by omiting the time $t$. This is very useful for calculating the dependence of $P_{n}$ on the intensity of the laser field as is shown in Fig. 4.    

\section*{Acknowledgments}
\vskip-0.6cm
We acknowledge support by the Greek funding program NSRF and the European Union's Seventh Framework Program FP7-REGPOT-2012-2013-1 under grant agreement 316165.

\section*{Author Contributions}
\vskip-0.6cm
I.A.G. obtained the closed-form solution of the TDSE, contributed on the quantum-optical description of the HHG and manuscript preparation; N.T. performed the theoretical calculations shown in the figures and contributed on the data analysis; I.K.K. contributed on the quantum-optical description of the HHG and manuscript preparation; P.T. conceived the idea and contributed in all aspects of the present work except of solving the TDSE.


\section*{References}
\vskip-0.6cm
\begin{thebibliography}{99}
	
	\bibitem{Keldysh}
	Keldysh, L. V. Ionization in the field of a strong electromagnetic wave. \textit{Sov. Phys. JETP} {\bf 20}, 1307 (1964).
	
	\bibitem{reiss}
	Reiss, H. R., Effect of an intense electromagnetic field on a weakly bound system. \textit{Phys. Rev. A} {\bf 22}, 1786 (1980).
	
	\bibitem{Krausz}
	Corkum, P. B., and Krausz, F., Attosecond science. \textit{Nat. Phys.} {\bf 3}, 381 (2007).
	
	\bibitem{Lew}
	Lewenstein, M., Balcou, Ph., Ivanov, M. Yu., L'Huillier, A., and Corkum, P. B., Theory of high-harmonic generation by low-frequency laser fields. \textit{Phys. Rev. A} {\bf 49}, 2117 (1994).
	
	\bibitem{Gao}
	Gao, J., Shen, F., Eden, J. G., Quantum electrodynamic treatment of harmonic generation in intense optical fields. \textit{Phys. Rev. Lett.} {\bf 81}, 1833 (1998).
	
	\bibitem{Gao3}
	Chen, J., Chen, S. G., and Liu, J., Comment on “Quantum Electrodynamic Treatment of Harmonic Generation in Intense Optical Fields”. \textit{Phys. Rev. Lett.} {\bf 84}, 4252, (2000).
	
	\bibitem{gaoguo1q2}
	Gao, J., Shen, F., Eden, J. G., Interpretation of high-order harmonic generation in terms of transitions between quantum Volkov states. \textit{Phys. Rev. A} {\bf 61}, 043812 (2000).
	
	\bibitem{Gao4}
	Hu, H., and Yuan, J., Time-dependent QED model for high-order harmonic generation in ultrashort intense laser pulses. \textit{Phys. Rev. A} {\bf 78}, 063826 (2008).
	
	\bibitem{quant}
	Mandel, L., and Wolf, E., Optical Coherence and Quantum Optics, Cambridge University Press, Cambridge, (1995).
	
	\bibitem{quant2}
	Schleich, W. P., Quantum optics in phase space, (John Wiley \& Sons, 2001).
	
	\bibitem{homodyne}
	Bachov, H. A., and Ralph, T. C., A guide to experiment in quantum optics, \textit{Wiley-VCH Verlag GmbH and Co.KGaA, Veinheim} (2004).
	
	\bibitem{Mlynek}
	Breitenbach, G., Schiller, S., and Mlynek, J., Measurement of the qauntum states of squeezed light, \textit{Nature} {\bf 387}, 471 (1997).
	
	\bibitem{Bellini}
	Zavatta, A., Parigi, V., Kim, M. S., and Bellini, M., Subtracting photons from arbitrary light fields: experimental test of coherent state invariance by single-photon annihilation, \textit{New. J. Phys.} {\bf 10}, 123006 (2008).
	
	\bibitem{bell}
	Bellini, M., \textit{et al.} Temporal Coherence of Ultrashort High-Order Harmonic Pulses. \textit{Phys. Rev. Lett.} {\bf 81}, 297, (1998).
	
	\bibitem{corsi}
	Corsi, C., Pirri, A., Sali, E., Tortora, A., and Bellini, M., Direct Interferometric Measurement of the Atomic Dipole Phase in High-Order Harmonic Generation. \textit{Phys. Rev. Lett.} {\bf 97}, 023901 (2006).
	
	\bibitem{zair}
	Za\"ir, A., \textit{et al.} Quantum Path Interferences in High-Order Harmonic Generation. \textit{Phys. Rev. Lett.} {\bf 100}, 143902 (2008).
	
	\bibitem{Kom}
	Kominis, I. K., Kolliopoulos, G., Charalambidis D., and Tzallas, P., Quantum-optical nature of the recollision process in high-order-harmonic generation. \textit{Phys. Rev. A} {\bf 89}, 063827 (2014).
	
	\bibitem{knight}
	Gerry, C., and Knight, P., Introductory Quantum Optics, Cambridge University Press, Cambridge, (2005).
	
	\bibitem{Hergott}
	Hergott, J.-F., \textit{et al.} Extreme-ultraviolet high-order harmonic pulses in the microjoule range. \textit{Phys. Rev. A} {\bf 66}, 021801(R) (2002).
	
	\bibitem{Tzallas1}
	Tzallas, P., \textit{et al.}, Generation of intense continuum extreme-ultraviolet radiation by many-cycle laser fields. \textit{Nat. Phys.} {\bf 3}, 846 (2007).
	
	\bibitem{Skantzakis}
	Skantzakis, E., Tzallas, P., Kruse, J., Kalpouzos, C., and Charalambidis, D., Coherent continuum extreme ultraviolet radiation in the sub-100nJ range generated by a high-power many-cycle laser field. \textit{Opt. Lett.} {\bf 34}, 1732 (2009).
	
	\bibitem{Tzallas2}
	Tzallas, P., Skantzakis, E., Nikolopoulos, L. A. A., Tsakiris, G. D., and Charalambidis, D., Extreme-ultraviolet pump-probe studies of one-femtosecond-scale electron dynamics. \textit{Nat. Phys.} {\bf 7}, 781 (2011).
	
	\bibitem{Mido}
	Takahashi, E. J., Lan, P., M\"{u}cke, O. D., Nabekawa, Y., and Modorikawa, K., Attosecond nonlinear optics using gigawatt-scale isolated attosecond pulses. \textit{Nat. Comm.} {\bf 4}, 2691 (2013).
	
	\bibitem{varro1q}
	Bergou, J., Varr\'o, S., Nonlinear scattering processes in the presence of a quantised radiation field. I. Non-relativistic treatment. \textit{J. Phys. A}: Math. Gen. {\bf 14}, 1469, (1981).
	
	\bibitem{my1q}
	Gonoskov, I. A., Vugalter, G. A., Mironov, V. A., Ionization in a Quantized Electromagnetic Field. \textit{J. Exp. Theor. Phys.} {\bf 105}, 1119, (2007).
	
	
\end{thebibliography}
\end{document}